\documentstyle [12pt]{article}
\begin{document}

\title {Polarized Beam Jets}
\author {John Collins \\
The Pennsylvania State University \\
104 Davey Laboratory \\
University Park PA 16802, U.S.A.}

\date{October 10, 1995}

\maketitle

\begin{abstract}
Handedness and a left-right asymmetry have been
proposed as methods of probing the polarization of hard
jets.  If the analyzing powers for these observables are
substantial they would provide useful tools for probing
polarized hard scattering.  This paper proposes that
comparing these to the corresponding quantities in beam
jets in polarized hadron-hadron collisions would be
useful.  These would shed light on the similarities and
differences between the fragmentation of hard and soft
jets.  They would also suggest whether there is any
useful possibility of measuring jet polarization.  The
left-right asymmetry of beam jets has already been
measured and is large.  But measurements of the
handedness of beam jets have still to be done.  They
could be done in an upgraded HERMES detector at HERA.
\end{abstract}


\section{Introduction}

Observables have been defined that enable one to probe
the spin of a quark initiating a jet.  These are
handedness \cite{Nacht,Efr78,EMT}, which provides a
measure of quark helicity, and the shared jet effect
\cite{trfrag,trfr-sgl,AC}, which is really a kind of
left-right asymmetry and which measures a quark's
transverse spin.\footnote{Many of these ideas can be
   found in an article by Baldracchini et al.\cite{Bald}
}
Such observables would be very useful for an experimental
program of polarized hadron collisions at high energy, such
has been proposed \cite{RSC} for the Relativistic Heavy Ion
Collider (RHIC) at Brookhaven---see, for example, the
papers by Stratmann and Vogelsang \cite{SV} and Nowak
\cite{Nowak}.

Now, these observables are proportional to quark spin,
but each with a constant  of proportionality, an
analyzing power, that is a property of the hadronization
process in the jet.  Unfortunately, these analyzing
powers are non-perturbative quantities that cannot at
present be predicted  from QCD.  Therefore it is highly
desirable to measure the analyzing powers in advance of
the proposed program of spin physics at RHIC;  if either
of them should happen to be substantial, there would be
a substantial impact on the direction of experimental
investigations.

The purpose of this paper is to propose some additional
ways of gaining insight into the sizes of these analyzing
powers, and in particular into their relative sizes.  I
observe that the bulk of events in a high energy
hadron-hadron collisions are kinematically two jet events.
Then the sheared jet and handedness observables can be
defined in identically the same manner as for the jets
resulting from a hard collision.  Aside from their
impact on the RHIC spin physics program, these measurements
have an intrinsic interest as a probe of the mechanism by
which soft beam jets are made.

So far, experimental observations have been made of the
transverse spin observable in soft collisions
\cite{E704,ssa} and of handedness \cite{SLD} in $e^{+}e^{-}$
annihilation at the $Z^{0}$; the first gives a large result
(tens of percent at large $x_{F}$) while the second has
resulted in an upper bound of a few percent.  The
transverse spin effect in hard jets can be measured
from jet-jet correlations in $e^{+}e^{-}$ annihilation, as
explained by Artru and Collins \cite{AC}.

One purpose of the present paper is to complete the set
of observables by suggesting measurement of the
handedness of a spectator jet in a collision of a
longitudinally polarized proton.   Since there appear to
be no suitable proton-proton experiments  available, a
second purpose is to show that the measurements can
be made in the target fragmentation region of polarized
electroproduction  experiments (in particular at low $Q^{2}$).
The HERMES \cite{HERMES} experiment appears
particularly suitable.

Finally, I will comment on the implications of these
measurements.  If soft and hard jets are similar as
regards their non-perturbative properties then one
predicts a large sheared jet effect in hard jets, but one
predicts a small handedness in soft jets.  If these
predictions are not verified by experiment then one has
a clear demonstration of large differences in the
non-perturbative hadronization mechanisms in the two
kinds of jets.

\section{Definition of Sheared Jet and Handedness Observables}

\subsection{Transverse Spin}

Let us consider a transversely polarized jet.  A hard jet in
this case would result from a quark of transverse
polarization, ${\bf s} _{\perp }$, that emanates from a hard
collision. A soft jet would be the beam jet in the collision
of a transversely polarized proton of transverse spin
${\bf s}_{\perp }$ with some other particle.  (We define a fully
polarized proton to have $|{\bf s}_{\perp }|=1$).

In either case, let ${\bf t}$ be the jet axis and consider
the distribution of hadrons in the jet as a function of
their momentum ${\bf p}$, in particular as a function of the
azimuthal angle $\phi $ of ${\bf p}$ about the jet axis. We have
a distribution proportional to
\begin{equation}
   1 + |{\bf s}|{\cal A}\cos\phi
   = 1 + \frac {{\bf s}\cdot  ({\bf t}\times {\bf p})}{|{\bf p}_{\perp }|}{\cal
A},
\end{equation}
where ${\cal A}(|{\bf p}_{\parallel}|,|{\bf p}_{\perp }|)$ is
the left-right asymmetry (or analyzing power) and ${\bf p}_{\perp }$
is the component of ${\bf p}$ perpendicular to the jet axis:
${\bf p}_{\perp }={\bf p}-{\bf t}\,({\bf t}\cdot {\bf p})$, where
we have assumed that ${\bf t}$ is a unit vector.  The
analyzing power ${\cal A}$ has a kinematic zero at
${\bf p}_{\perp } = 0$.

The necessity for the above form of asymmetry arises from
the facts that spin dependence of a cross-section is linear
in ${\bf s}$ and that in a parity conserving theory the only
scalar quantity we can construct is ${\bf s} \cdot ({\bf t} \times {\bf p})$.

In the case of a soft beam jet, ${\bf t}$ is the direction of the
incoming polarized proton, and the usual notation is to use
$A_{N}$ instead of ${\cal A}$.  Measurements have been made
\cite{E704,ssa} as function of $x_{F}$, and reach values of
30\% to 40\% for $x_{F} > 0.7$, but are small for $x_{F} < 0.3$,
in the case of charged pions, and somewhat smaller for $\pi ^{0}$.

Collins \cite{trfrag,trfr-sgl} defined a `sheared jet
asymmetry' by exactly the same formula, but applied it
to a polarized hard jet, such as would result in deep
inelastic lepton scattering (DIS) on a transversely
polarized proton.  The variable $x_{F}$ new gets replaced by
$z$---the fractional momentum of the measured hadron in
the jet.  In the DIS case, the jet axis ${\bf t}$ can be defined
(in the center-of-mass frame) by parton-model
kinematics.

In a more general case one could use a standard jet algorithm to
define ${\bf t}$.  Alternatively, one can use the 2 leading particles
in the jet to define both the jet axis and the azimuthal angle; one
replaces ${\bf s} \cdot ({\bf t} \times {\bf p}) / |p_{\perp }|$ by
\begin{equation}
   \frac {{\bf s}\cdot ({\bf p}_{1}\times {\bf p}_{2})}{ |{\bf p}_{1}\times
{\bf p}_{2}|} .
\end{equation}

\subsection{Longitudinal Spin}

Let us now replace the transverse polarization of the
initiator of a jet by a longitudinal polarization (or helicity $\lambda $).
Then to get a scalar quantity linear in the pseudoscalar $\lambda $ we
need an additional momentum.

One possibility is to have a jet axis and two measured particles:
\begin{equation}
   \lambda {\bf t}\cdot ({\bf p}_{1}\times {\bf p}_{2}) .
\end{equation}
This is what is used in the SLD measurement of
handedness \cite{SLD}.

A second possibility, the one original given by
Nachtmann \cite{Nacht}, is to measure three particles in
the final state and then we have a possible helicity
dependence of the form
\begin{equation}
   \lambda {\bf p}_{1}\cdot ({\bf p}_{2}\times {\bf p}_{3}) .
\end{equation}
A convenient way to perform a measurement of the dependence
of the cross-section on one of these variables is to define
\begin{equation}
   \Omega ={\bf t}\cdot ({\bf p}_{1}\times {\bf p}_{2})\label{omega}
\end{equation}
for the 2 particle plus jet measurement.   Then we define handedness,
H, as the asymmetry between the number of events with positive and
negative $\Omega $:
\begin{equation}
   H=\frac {N(\Omega >0)-N(\Omega <0)}{N(\Omega >0)+ N(\Omega <0)} ,
\label{H}
\end{equation}
where the numbers of events are computed with some suitable
cuts on the particles.  For example, a minimum $p_{\perp }$
prevents the value of $H$ from being diluted by the kinematic
zero at zero $p_{\perp }$, and a minimum $x_{F}$ or $z$
forces the hadrons to be in the `valence region' where the spin
dependence is presumably largest.  The results should be
presented for $H/ \lambda $.

The SLD collaboration measured the handedness of jets
produced in $e^{+}e^{-}$ annihilation at the $Z^{0}$; where quarks
and antiquarks are produced with known, large helicities.
The experimenters had to distinguish, in effect, quark and
antiquark jets, since their helicities are opposite.

\section{Handedness in Soft Jets}

\subsection{$pp$ and $ep$ experiments}

I propose that the handedness of a beam jet be measured in inclusive
2 pion production in relativistic collisions of
\begin{equation}
   \vec{p} p \rightarrow \pi ^{+}\pi ^{-}X .
\label{pp}
\end{equation}
Handedness, as defined by Eqs.\ (\ref{omega}) and (\ref{H}),
is proportional to the helicity $\lambda $ of the incoming proton;
it will reverse sign when $\lambda $ is reversed.  In Eq.\
(\ref{omega}) we let ${\bf p}_{1}$ and ${\bf p}_{2}$ be the
momenta of the $\pi ^{+}$ and $\pi ^{-}$.

Appropriate minima on the values of the transverse and
longitudinal momenta of the 2 pions should be imposed.
Appropriate values might be
$x_{F1}, x_{F2} > 0.3$, $p_{1\perp }, p_{2\perp } > 0.3\ {\rm GeV}$.
The only recent experiment that could make the measurement is
E704 \cite{E704} at Fermilab, but its coverage in azimuth appears
to be insufficient to make an analysis of the data worthwhile.

However, the HERMES \cite{HERMES} experiment at DESY
is appropriate.  It performs collisions of 30 GeV
electrons on a polarized gas jet target, so we replace
the reaction (\ref{pp}) by
\begin{equation}
   \gamma ^{*}\vec{p} \rightarrow \pi ^{+}\pi ^{-}X, \mbox{\rm \ (or
equivalently}\;
   e\vec{p}  \rightarrow e\pi ^{+}\pi ^{-}X) ,
\end{equation}
where now we require the measured pions to be in the target
fragmentation region.  The virtuality of the $\gamma $ is
irrelevant --- indeed the closest analogy with the
$\vec{p}p$ reaction is at low $Q^{2}$ rather than in the
deep-inelastic region. Furthermore, to get into the 2-jet
region, one should require the hadronic final state to have
a high mass --- particularly in view of the low energy of
the beam.  Any polarization of the virtual photon (or of the
electron producing it) is irrelevant for our purposes.

Of course, it would also be of interest to measure
handedness both for deep-inelastic events and for lower mass
final states.

It is desirable to strengthen the cuts on the measured pions
to maximize the measured handedness.  The need for this is
clearly  indicated by the way in which the left-right
asymmetry in the case of transverse spin rises strongly with
$x_{F}$ and with $p_{\perp }$.  But one must also make measurements
with cuts that correspond to those used by the SLD
measurements \cite{SLD}.

\subsection{Other Measurements for Soft Jets}

To verify that the proton jet in polarized electroproduction
behaves like the beam jet in polarized proton-proton scattering,
one needs the single transverse spin asymmetry, i.e. the
left-right asymmetry in
\begin{equation}
   \gamma ^{*}p^{\uparrow} \rightarrow \pi X .
\end{equation}
Here the proton is transversely polarized and the pion is in the
target fragmentation region. It would require very
implausible dynamics for the left-right
asymmetry if the results were greatly different from those in
$pp$ scattering.  In this and other reactions in this paper,
we use $p^{\uparrow}$, with an upward pointing arrow, to
denote a transversely polarized proton, and $\vec {p}$ to denote
a longitudinally polarized proton.

In addition, measurements should be made of a two-particle
asymmetry with a transversely polarized proton:
\begin{equation}
   \gamma ^{*}p^{\uparrow} \rightarrow \pi ^{+}\pi ^{-}X .
\end{equation}
The measurement is of the coefficient ${\cal A}$ in
\begin{equation}
   1 + {\cal A}
      \frac {{\bf s}\cdot ({\bf p}_{1}\times {\bf p}_{ 2})}{|{\bf p}_{1}\times
{\bf p}_{2}|} ,
\end{equation}
where the momentum of one of the measured pions is used
instead of the beam axis to define a left-right asymmetry.
One should compare this two-particle asymmetry with the
conventional one-particle left-right asymmetry.  Any dilution
 of the asymmetry that occurs here by adding an extra particle
 would likely also occur for hard polarized jets.
Compare \cite{trfrag} and \cite{trfr-sgl} for the
corresponding definitions for hard jets.

\section{Predictions}

A hard jet arises when a fast moving quark (or gluon) is created
in a short-distance process.  The early history of the jet involves
quark and gluon radiation governed by perturbative dynamics.
The late non-perturbative stages involve an expanded and
elongated clump of hadronic matter.  In contrast, a soft jet is
created by an already extended clump of hadronic matter
moving away at high rapidity from a collision.

Thus although the early history of the two kinds of jet is different,
it is reasonable to suppose that the non-perturbative part of their
 hadronization is qualitatively similar, and in particular that the spin
 effects are similar.  On the basis of existing measurements, I
predict
\begin{itemize}
\item Small handedness (at most a few \%) in soft jets.
\item Large left-right asymmetries (tens of \%) in hard jets at
$z>$ 0.3 to 0.5.
\end{itemize}

\noindent
If this is so, then a nonzero sheared-jet effect should be measurable
from current $e^{+}e^{-}$ data.  This is not an absolute
prediction of QCD, since we do not understand its non-perturbative
dynamics very well.  If, on the contrary, spin
effects are substantially different between the two kinds of
 jets then that has big implications for models of fragmentation.
(There is one obvious difference between soft jets and hard
jets at large Q.  This is caused by Altarelli-Parisi evolution.
Its effects can be minimized by restricting one's attention
to clean low multiplicity jets, i.e., ones with little radiation.)

\section*{Acknowledgments}

My work was supported in part by the U.S. Department of
Energy under grant number DE-FG02-90ER-40577.  I would like
to thank colleagues for discussions, notably X. Artru and M.
Strikman.


\end{document}